# Refining Gaussian Splatting: A Volumetric Densification Approach

Mohamed ABDUL GAFOOR     Marius PREDA     Titus ZAHARIA

Laboratoire SAMOVAR, Telecom SudParis, Institut Polytechnique de Paris  
9 Rue Charles Fourier 91000, Courcouronnes, France

mohamed.abdul_gafoor@telecom-sudparis.eu     marius.preda@telecom-sudparis.eu     titus.zaharia@telecom-sudparis.eu

## ABSTRACT

Achieving high-quality novel view synthesis in 3D Gaussian Splatting (3DGS) often depends on effective point primitive management. The underlying Adaptive Density Control (ADC) process addresses this issue by automating densification and pruning. Yet, the vanilla 3DGS densification strategy shows key shortcomings. To address this issue, in this paper we introduce a novel density control method, which exploits the volumes of inertia associated to each Gaussian function to guide the refinement process. Furthermore, we study the effect of both traditional Structure from Motion (SfM) and Deep Image Matching (DIM) methods for point cloud initialization. Extensive experimental evaluations on the Mip-NeRF 360 dataset demonstrate that our approach surpasses 3DGS in reconstruction quality, delivering encouraging performance across diverse scenes.

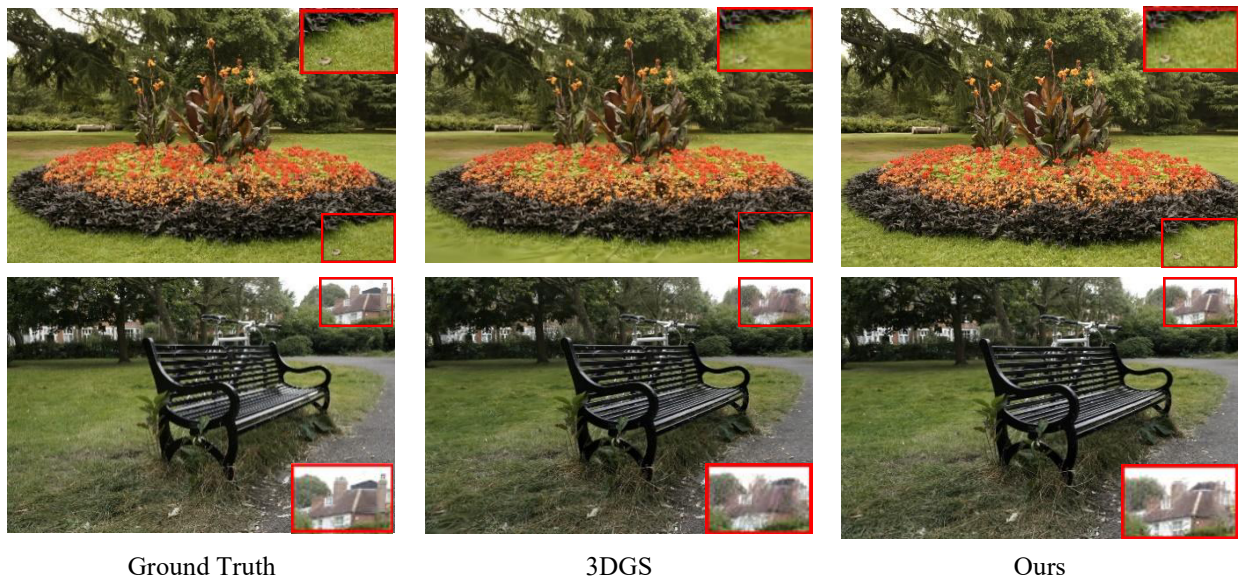

Figure 1: Challenges faced by 3DGS: the method fails to render some specific regions, where the related density of Gaussians is insufficient, leading to sparse details and blur. By modifying the densification process, the proposed approach overcome such shortcomings, delivering accurate reconstructions.

## 1. Introduction

High-quality Novel View Synthesis (NVS) is a fundamental task in computer vision and computer graphics, with diverse applications in AR/VR/MR, robotics, and cinematography. The objective is to generate photorealistic images of a scene from unseen viewpoints, given a set of input images with known camera parameters.

Early methods such as Structure-from-Motion (SfM) [1] and image-based rendering (IBR) have gradually evolved towards more sophisticated neural representations. A pivotal breakthrough came with the Neural Radiance Fields (NeRF) approach [2], which leverages a multi-layer perceptron (MLP) neural network to represent scenes as continuous fields of color and density. NeRF-based methods optimize scene representations using volumetric rendering and have demonstrated remarkable image quality, capturing intricate lighting and geometric effects. However, they suffer from slow training and inference, which limits their applicability in real-time scenarios [2], [3]. In response to such limitations, 3D Gaussian Splatting (3DGS) [4] has emerged as a promising alternative. Unlike NeRF's implicit representation via MLPs, 3DGS uses an explicit set of 3D Gaussian primitives. These Gaussians are characterized by position, covariance, opacity, and color attributes. Splatting-based rasterization is then employed to project Gaussians onto 2D images from various angles of view, enabling high-resolution real-time rendering. Each Gaussian's appearance is modeled with anisotropic spherical harmonics (SH) [5]. Importantly, 3DGS avoids computationally expensive ray marching and point sampling, making it well-suited for fast rendering [4]. One of the challenges in 3DGS concerns the optimization of the *Adaptive Density Control* (ADC) mechanism [4], whose role is to adapt the density of Gaussians to the specificity of each image region. In the vanilla formulation, this is done by either inserting or pruning Gaussian primitives based on gradients and opacity thresholds. However, the approach fails in certain cases, leading to sparse details and blur (Fig. 1). In this paper, we specifically tackle this issue and propose the following contributions:

1. A complementary densification mechanism, that automatically identifies over-sized Gaussians based on volumetric criteria and facilitates the generation of new ones whenever necessary.

2. An experimental evaluation of two different initialization procedures, which include the traditional SfM process [1] and a Deep-Image-Matching (DIM) framework [6], [7], [8] with *SuperPoint* [9] as feature extractor and *LightGlue* [10] as feature matcher.

The goal here is to identify the cases for which each initialization technique is more appropriate.

The rest of the paper is organized as follows. Section 2 reviews the related work on 3D reconstruction techniques, and in particular focuses on research dedicated to improvements of 3DGS-related methodologies. Section 3 recalls the 3DGS mathematical model and introduces the related notations. The proposed approach is described in details in Section 4. The experimental setup and evaluation are presented and discussed in Section 5. Finally, Section 6 concludes the paper and opens some perspectives of future work.

## 2. 3D Reconstruction: related work

Traditional 3D reconstruction methods are based on well-established approaches such as Structure from Motion (SfM) and Multi-View Stereo (MVS) [1], [11]. However, their limitations in terms of robustness, efficiency, and scalability make them less competitive compared to modern deep learning based methods.

In 2015, Sergey *et al.* [12] propose a novel approach to learn a general similarity function for comparing image patches using Convolutional Neural Networks (CNNs). Unlike traditional methods that rely on handcrafted features like SIFT, the features are here learned directly from raw image data. This work has made a significant contribution to the field, demonstrating how deep learning can surpass traditional feature descriptors.

In [13], Kwang *et al.* introduce an end-to-end pipeline combining keypoint detection, descriptor extraction and orientation estimation within a unified pipeline. Subsequently, the so-called *SuperPoint* [9], *SuperGlue* [14] and *LightGlue* [10] techniques propose further optimizations, enhancing the computational efficiency while maintaining robustness to viewpoint changes, or scale variations.

The Scene Representation Networks (SRNs) introduced by Sitzmann *et al.* [15] exploits a continuous, 3D structure-aware neural scene representation that encodes both geometry and appearance without explicit 3D supervision. In 2020, Mildenhall *et al.* introduced the NeRF (*Neural Radiance Fields*) model [2]. The method makes it possible to generate views of complex scenes by using a continuous volumetric scene function and a small set of input images. However, the related limitations concern the high computational complexity, which notably leads to excessively slow training times. More recently, in 2023, the 3D Gaussian Splatting (3DGS) method has revolutionized the field by proposing an efficient model for representing and rendering volumetric data. The principle consists in modeling scenes with a set of anisotropic Gaussian primitives [4]. Numerous applications leveraging 3DGS have been proposed, including Gaussian-based avatars for realistic human representation [16], [17], [18], autonomous driving for efficient scene understanding [19], [20], [21] and dynamic scene reconstruction to capture temporal changes [22], [23], [24]. Recent advances focus on improving its scalability, fidelity, and integration with neural radiance fields to enhance the scene representation [25].

Several studies emphasize the importance of addressing fundamental improvements in 3D Gaussian Splatting methodologies such as rendering quality, efficiency and memory optimization. Yu *et al.* [26] introduce a 3D smoothing filter to regularize the frequency of 3D Gaussian primitives and replace the 2D dilation filter with a 2D Mip filter to address aliasing and dilation artifacts, thus improving rendering quality and handling out-of-distribution issues. Lu *et al.* [27] propose a hierarchical and region-aware scene representation by initializing anchor points from a sparse voxel grid to guide the distribution of local 3D Gaussians. They further predict neural Gaussians on-the-fly within the view frustum to accommodate varying perspectives and distances, enabling robust novel view synthesis.

To overcome the limitations of spherical harmonics in modeling specular and anisotropic components, Yang *et al.* [28] introduced anisotropic spherical Gaussians and a coarse-to-fine training strategy, significantly enhancing the rendering quality for complex surfaces without increasing the underlying Gaussian counts. Lee *et al.* [29] proposes a learnable masking mechanism based on the volume and opacity of Gaussians. Hence, Gaussians with small volumes or low opacity are identified as redundant and removed using binary masks. The GaussianPro approach [30] addresses densification challenges related to the shortcomings of SfM-based initialization. Here, a sophisticated approach progressively propagates primitives along estimated planes, guided by patch-matching techniques and geometric consistency criteria. Huang *et al.* [31] analyze the correlations between rendering errors and Gaussian positions. A mathematical estimation of such errors is derived, which makes it possible to determine its extrema using optimization techniques. Bulò *et al.* [32] enhances 3DGS's ADC mechanism by guiding densification using a per-pixel error function instead of positional gradients. In addition, the influence of biases in the Gaussian cloning process is also addressed.

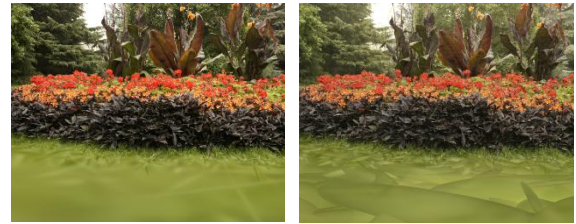

Figure 2: The 3DGS rendered image with a blurry grass region (left) and its corresponding overlay of ellipsoids (right).

However, none of the above methods address the need of densification in regions where the Gaussian cloud remains excessively sparse. Such a case is illustrated in Fig. 2. Here, the grass region presents a relatively moderate texture that is rendered in a blurry manner by 3DGS (left). The corresponding Ellipsoids (right) remain in this case over-sized.

In this paper we notably identify and tackle this issue by introducing a complementary, volume-driven Gaussian densification approach.

## 3. Gaussian splatting: problem statement

In Gaussian splatting, a 3D scene is represented as a collection of $N$ anisotropic 3D Gaussian primitives [4], denoted by $\Gamma = \{\gamma_1, \gamma_2, \ldots \gamma_N\}$.

Each primitive $\gamma_i = (\mu_i, \Sigma_i, \alpha_i, f_i)$ is defined by a 3D Gaussian kernel:

$$G_i(x) = \exp\left(-\frac{1}{2}(x - \mu_i)^T \Sigma_i^{-1}(x - \mu_i)\right), \quad (1)$$

where $\mu_i \in \mathbb{R}^3$ is represents the Gaussian's center and $\Sigma_i \in \mathbb{R}^{3\times 3}$ its covariance matrix, shaping its spatial extent and orientation. To enforce the covariance matrix to remain positive semi-definite during the optimization process, $\Sigma_i$ is expressed as:

$$\Sigma_i = R_i S_i^2 R_i^T \quad (2)$$

with $R_i$ being an orthogonal rotation matrix and $S_i$ a diagonal scale matrix [30]. Each Gaussian is further enriched with an opacity factor $\alpha_i \in [0,1]$ and a feature vector $f_i \in R^d$, which may include attributes such as RGB color or spherical harmonics coefficients.

The 3D Gaussian primitives are rendered using splatting-based rasterization. This process involves the projection of a 3D Gaussian primitive $\gamma_i$ from the 3D world onto the 2D image space using a transformation $\pi: \mathbb{R}^3 \to \mathbb{R}^2$. To simplify the computation, the projection is locally linearized at the center $\mu_i$ of the Gaussian. In this way, the primitive is approximated as a 2D Gaussian $G_i^\pi$ in the image space, where the mean of the projected Gaussian becomes $\pi(\mu_i) \in \mathbb{R}^2$, and the 2D covariance matrix of the projected Gaussian is defined as:

$$\Sigma_i^\pi = J_i^\pi \Sigma_i J_i^{\pi T} \qquad (3)$$

where $\Sigma_i^\pi$ represents the covariance matrix in image space, and $J_i^\pi$ is the Jacobian of the transformation $\pi$, evaluated at $\mu_i$ [32].

The color $c(p)$ of each pixel $p$ is determined by blending $N$ Gaussians $\Gamma = \{\gamma_1, \gamma_2, \ldots \gamma_N\}$, that overlap the given pixel. The blending process is defined as:

$$c(p) = \sum_{i=1}^{N} c_i \alpha_i \prod_{j=1}^{i-1}(1 - \alpha_j) \qquad (4)$$

where $\alpha_i$ represents the opacity of the Gaussian $G_i^\pi$, computed as the product between $G_i^\pi(p)$ and a learned opacity parameter associated to $\gamma_i$ [4]. The color $c_i$ is also a learnable attribute of $\gamma_i$. The Gaussians overlapping at pixel $p$ are sorted in ascending order based on their depth relative to the current viewpoint to ensure proper compositing. Using differentiable rendering techniques, all Gaussian attributes, including position, shape, color, and opacity, can be optimized in an end-to-end manner through training view reconstruction [4].

Furthermore, during an additional optimization process, the ADC mechanism allows the dynamic addition and removal of 3D Gaussians to enhance the scene representation [4].

## 4. Proposed volume-based densification

The 3DGS construction process requires an initialization, which is traditionally achieved with SfM techniques. However, such procedures fail to generate a sufficient number of 3D points in relatively poorly textured regions, such as grass or sand, resulting in gaps in the reconstruction. This weak initialization affects the ADC mechanism, making it difficult to produce reliable 3D Gaussians and leading to insufficient scene coverage [30]. To enhance the density of the initial point cloud, we have also considered as an alternative a DIM technique [6], an example is illustrated in Figure 3. While DIM improves here the Gaussians's density when compared to SfM, both methods still struggle to capture relatively low-textured regions effectively, which may penalize the reconstruction quality. To address this issue, we introduce a complementary mechanism to ADC that goes beyond splitting and cloning [4]. The principle consists of refining the 3D Gaussian generation process based on a supplementary volumetric criterion.

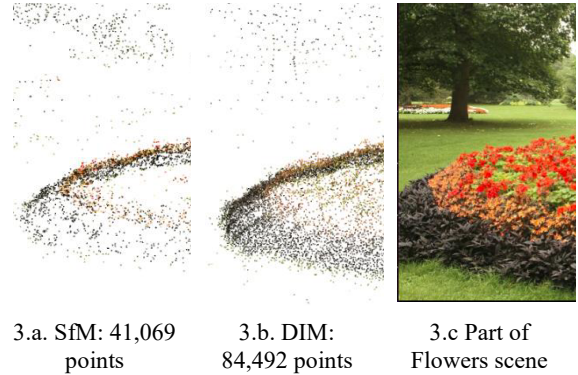

3.a. SfM: 41,069 points   3.b. DIM: 84,492 points   3.c Part of Flowers scene

Figure 3: Initial point clouds: SfM *versus* DIM for the Flowers scene (Mip-NeRF360 dataset).

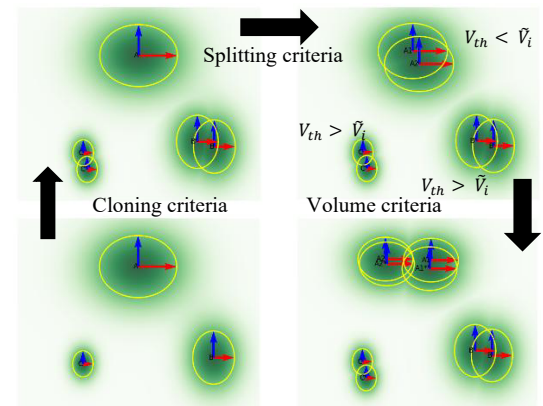

Figure 4: 3D Gaussians undergo gradient-based cloning and splitting as in [4], followed by the proposed volume-based densification process.

More precisely, for each Gaussian primitive, we consider its ellipsoid of inertia, whose axes are aligned with the eigenvectors of the Gaussian covariance matrix $\Sigma$ with semi-lengths defined by the corresponding eigenvalues $\sqrt{\lambda_1}, \sqrt{\lambda_2}, \sqrt{\lambda_3}$.

During training, we compute the volume of each Gaussian ellipsoid. The determinant of $\Sigma$ is equal to the product of its eigenvalues, $\det(\Sigma) = \lambda_1 \lambda_2 \lambda_3$. Hence the ellipsoid's volume can be calculated as:

$$V = \frac{4}{3}\pi\sqrt{\det(\Sigma)} \qquad (5)$$

If the volume of a given Gaussian ellipsoid $\tilde{V}_i$ is above a pre-defined threshold $V_{th}$, then the ellipsoid is considered for further densification through a split mechanism (Fig. 4).

Selecting an appropriate threshold is essential for achieving a good quality of the rendered images. Setting the volume threshold $V_{th}$ too low can result in the generation of new points in already dense regions. On the contrary, a too large value would fail to take into account regions that need to be further densified.

In addition, in order to adjust the standard deviation of the resulting splitted Gaussians, we calculate the condition number [33], defined as the ratio between the largest and the smallest eigenvalues of Σ:

$$\kappa(\Sigma) = \frac{\lambda_{max}(\Sigma)}{\lambda_{min}(\Sigma)} \quad (6)$$

The condition number provides a measure of the degree of elongation of the considered Gaussians. For isotropic Gaussians (*i.e.*, equal eigenvalues), its value equals 1. In contrast, anisotropic Gaussians, which present an elongated shape towards a large dominant direction, present a large condition number. During the splitting process, the eigenvalues of each newly generated Gaussian are adjusted as described in the following equation:

$$\forall\, i \in \{1, 2, 3\}, \quad \lambda'_i = \frac{\lambda_i}{\kappa(\Sigma)} \quad (7)$$

In the isotropic case, the size of the resulting ellipsoids does not change after split. In contrast, for the anisotropic one, the size is divided by the condition number (while their orientations and global aspect ratio are preserved). Similarly, to the cloning and splitting operation in the native 3DGS approach, the volume-based densification is applied iteratively every 100 iterations.

The proposed method makes it possible to densify sparsely-represented regions, where high volume ellipsoids of elongated aspect ratios may appear. By considering all relevant eigenvalues in the densification procedure, it ensures a comprehensive representation of the Gaussian spread in all directions.

## 5. Experimental evaluation

In our experiments, we have considered various samples from Mip-NeRF 360 [34], Tanks and Temples [35], and Deep Blending [36] datasets. For Mip-NeRF 360, we have included all scenes, comprising five outdoor scenes and four indoor scenes. From the Tanks and Temples dataset, we have selected two specific scenes: Dr. Johnson and Playroom. Similarly, for Deep Blending, we focused on two datasets: Truck and Train.

To evaluate performances, we have retained the three metrics usually considered in the state of the art: peak signal-to-noise ratio (PSNR), structural similarity index (SSIM), and the perceptual metric LPIPS [37].

We have adopted the original Gaussian Splatting implementation available in the GitHub repository (https://github.com/graphdeco-inria/gaussian-splatting) and extended it with the proposed volume-based densification procedure. All experiments have been run on a NVIDIA RTX 2080 GPU.

In our experiments, we found that the majority of ellipsoids have small volumes, with a rapid drop of the relative frequency of appearance as the volume increases. Fig. 5 illustrates the volume's histogram for the MiP-NeRF 360 dataset. We have collected volume data by running 3DGS [4] on the scenes at iterations 4000, 8000, and 12000 to generate a histogram plot, providing a means to visualize the volume of large ellipsoids. To balance efficiency and visual quality, we set a threshold of 0.03, as indicated in the inset (Fig. 5). Let us note that more than 99.78% of the ellipsoids' volumes fall below this threshold. So, this choice helps to densify solely the remaining large ellipsoids. However, this procedure can significantly enhance the quality of the rendered images, as illustrated in the example presented in Fig. 6. Here, in the grass region the ellipsoids generated by the 3DGS approach are excessively over-sized and consequently yield blurry images. By applying the volumetric-based densification procedure this limitation is overcome.

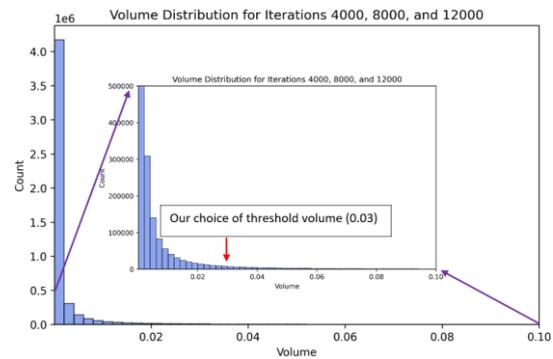

Figure 5: Histogram of the overall volume distribution for the Mip-NeRF 360 dataset, sampled at 4000, 8000, and 12000 iterations.

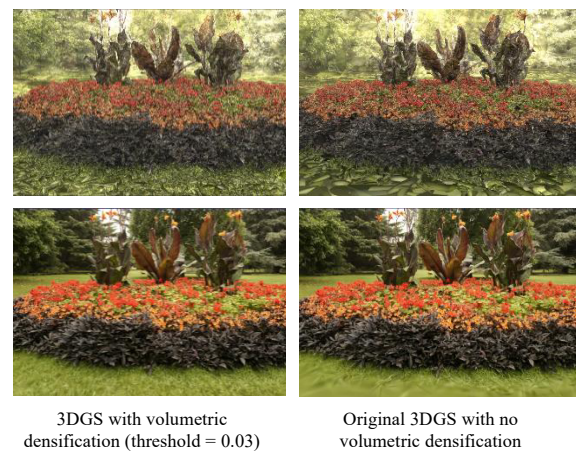

3DGS with volumetric densification (threshold = 0.03)    Original 3DGS with no volumetric densification

Figure 6: Effect of the volumetric densification (threshold $V_{th} = 0.03$).

| Dataset | SSIM | | PSNR | | LPIPS | | Gaussian (million) | Gaussian (million) | Train (min) | Train (min) | Scenes |
|---|---|---|---|---|---|---|---|---|---|---|---|
| | 3DGS | Ours | 3DGS | Ours | 3DGS | Ours | 3DGS | Ours | 3DGS | Ours | |
| Mip-NeRF 360 | 0.69 | 0.69 | 23.56 | 23.50 | 0.29 | **0.26** | 2.37 | 3.25 | 42.3 | 69.2 | Flowers |
| | 0.60 | 0.59 | 22.45 | 22.16 | 0.36 | **0.33** | 3.22 | 3.42 | 58.6 | 81.0 | Tree-hill |
| | 0.72 | **0.75** | 24.89 | **24.97** | 0.27 | **0.22** | 3.41 | 3.45 | 74.6 | 83.4 | Bicycle |
| | 0.85 | 0.85 | 27.06 | **27.21** | 0.13 | **0.12** | 3.41 | 3.41 | 76.1 | 79.5 | Garden |
| | 0.94 | **0.95** | 32.97 | **33.53** | 0.18 | **0.15** | 0.96 | 2.51 | 33.0 | 73.6 | Bonsai |
| | 0.90 | **0.91** | 29.01 | **29.21** | 0.21 | **0.19** | 1.08 | 1.27 | 41.5 | 48.8 | Counter |
| | 0.91 | **0.92** | 32.09 | **32.65** | 0.22 | **0.19** | 0.92 | 1.19 | 40.6 | 49.8 | Room |
| | 0.93 | 0.93 | 31.00 | **31.13** | 0.12 | 0.12 | 1.47 | 1.67 | 59.9 | 69.8 | Kitchen |
| | 0.77 | 0.74 | 26.61 | 25.96 | 0.23 | 0.23 | 3.43 | 3.57 | 62.3 | 87.3 | Stump |
| Tanks and Temples | 0.86 | **0.87** | 24.53 | **24.95** | 0.16 | **0.12** | 1.94 | 3.41 | 27.1 | 73.7 | Truck |
| | 0.58 | 0.57 | 16.60 | 16.47 | 0.40 | **0.38** | 0.97 | 3.15 | 32.3 | 78.9 | Train |
| Deep Blending | 0.86 | 0.85 | 27.83 | 27.70 | 0.31 | **0.29** | 2.35 | 3.35 | 57.3 | 86.9 | Playroom |
| | 0.74 | 0.74 | 20.76 | 20.65 | 0.44 | 0.44 | 1.77 | 1.93 | 49.8 | 55.5 | Dr.Johnson |

Table 1: Average results of three trials for both 3D Gaussian Splatting and proposed method (DIM initialization).

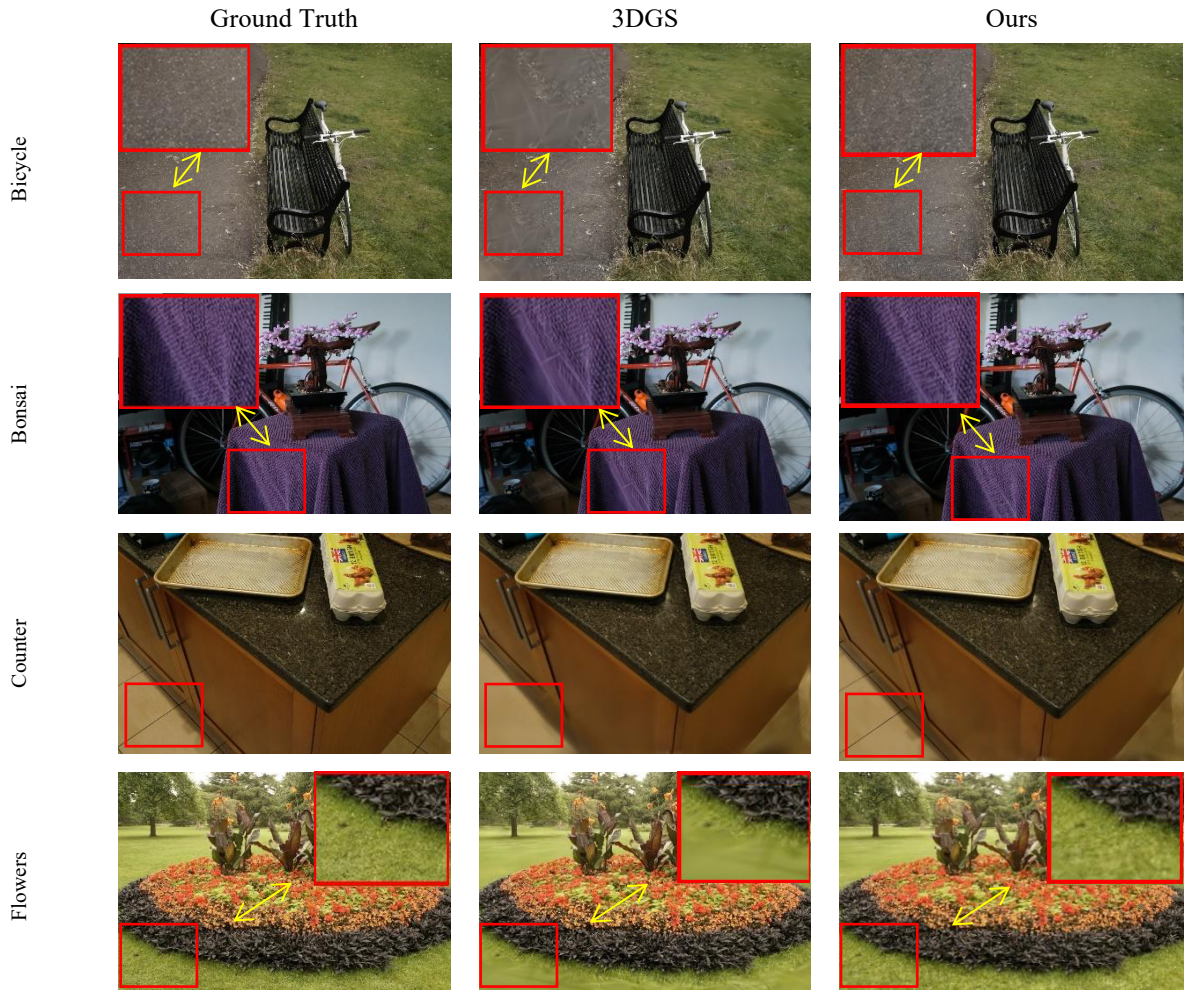

Figure 7: Qualitative comparison on MiP-NeRF 360 dataset.

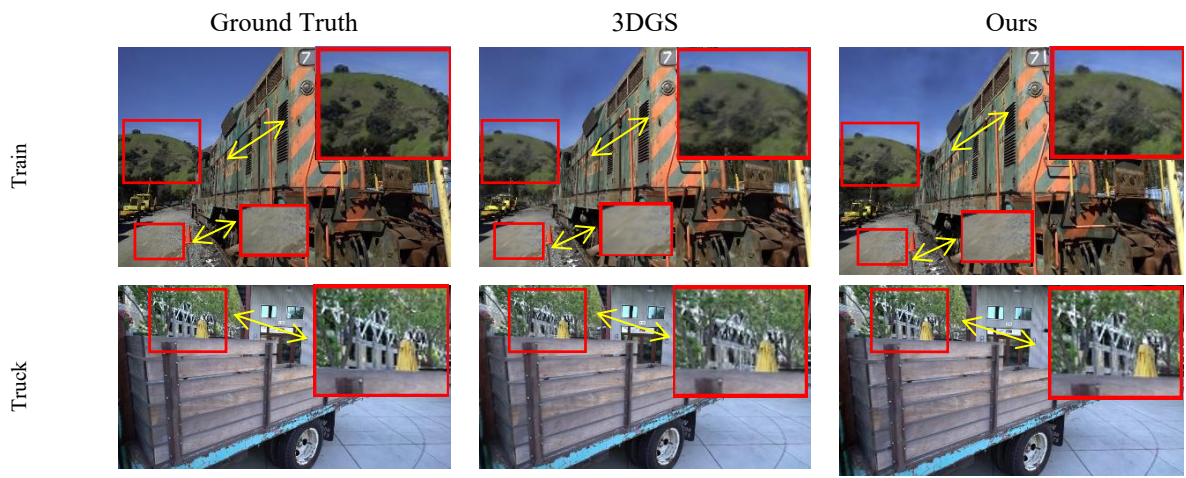

Figure 8: Qualitative comparison on Tanks and Temples dataset.

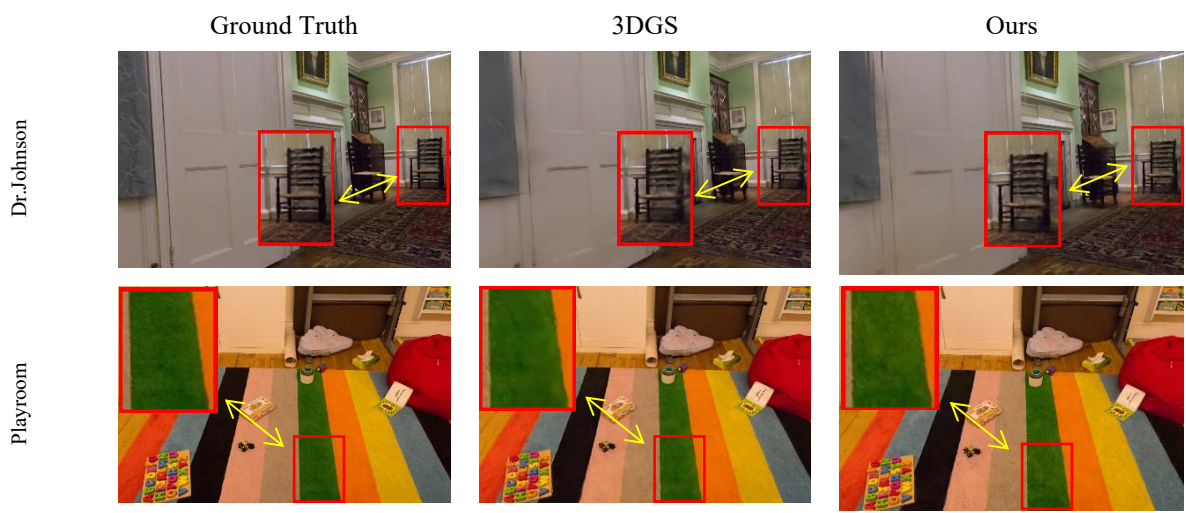

Figure 9: Qualitative comparison on Deep Blending dataset.

To manage computational resources, we have imposed a ceiling on the number of points generated during training, capping this value at 3,4 millions. This limit can be adjusted as needed, providing flexibility in handling different datasets. We have also restricted the number of images in each scene to a maximum value of 200 to avoid memory-related issues. Although the MiP-NeRF 360 dataset features image resolutions in the 2K, 3K and 5K ranges, we have opted for relatively homogeneous resolutions, ranging from 1237×822 to 1555×1039 pixels. In this way, we avoid out of memory issues. To this purpose, we have down-sampled the original images by ×2 or ×4 factors whenever necessary. In the same time, this makes it possible to define a unique volumetric threshold for all images considered, since the volume measure is not scale invariant. During training, we utilized the Adam optimizer with learning rates adopted from the 3DGS paper [4]. Additionally, for each scene in the dataset we run the algorithm three times and take the average values of PSNR, SSIM and LPIPS. The same procedure has been applied for the baseline 3DGS method that we have considered with the default configurations.

Table 1 presents a quantitative comparison between our method and the baseline 3DGS technique across the three datasets retained. A DIM initialization has been applied here as well as in all examples illustrated in Fig. 7 to 9. Our approach consistently achieves lower LPIPS scores, indicating better perceptual quality and closer resemblance to the ground truth. It also performs competitively in PSNR and, to a lesser extent, SSIM improvements can be observed. Concerning the number of Gaussians of the representation, our approach slightly augments it, since the method is specifically designed to this purpose. However, the increase remains reasonable, with a global average of 30.33% on the whole test dataset retained. Despite this augmentation, the method continues to deliver performance suited for real-time rendering applications for all the scenes in our dataset.

Figures 7, 8 and 9 present some results from the Mip-NeRF360, Tanks and Temples and Deep Blending datasets. The Bicycle, Bonsai and Flowers scenes in Fig. 7 demonstrate that our approach produces results significantly closer to the ground truth, with enhanced texture fidelity. In contrast, 3DGS introduces noticeable artifacts, particularly in ground regions (highlighted in red), where textures appear smoothed out or distorted. Our method effectively reconstructs fine details, preserving the texture and greatly improving realism. Likewise, in the Train scene (Fig. 8), our method shows an improvement in the ground region and mountain areas compared to 3DGS. Similarly, in the Playroom scene (Fig. 9), our method faithfully reconstructs the carpet areas, where texture details are over-smoothed by 3DGS. Another interesting feature of the proposed method concerns its ability to preserve directional image structures. This can be observed in the Counter scene (Fig. 7), where 3DGS completely eliminates an edge that is present on the floor pavement, whereas our method accurately preserves it. Similarly, in the Truck scene (Fig. 8) we can see edge information is preserved in the building structure. Maintaining sharp object edges is crucial for ensuring realistic reconstructions that can be useful for applications like AR/VR. Finally, in the case of the Dr. Johnson scene (Fig. 9), we do not observe any significant improvement, which is consistent with the metric reported in Table 1.

In a second stage, we have analyzed the impact of the initialization approach on the global performances. Table 2 presents the LPIPS values obtained by comparing the two different initialization strategies considered (SfM and DIM), when combined with both the baseline 3DGS and the proposed method across all scenes. The results indicate that, in a general manner, the proposed method achieves lower LPIPS values for both SfM and DIM, demonstrating here again its capacity of achieving a superior perceptual quality. The SfM-based initialization performs better in a majority of cases. However, it fails in the case of Flowers, Truck, and Bicycle scenes, where DIM-based initialization yields better results. This behavior can be explained by the fact that the DIM-based initialization is particularly effective for dense, repetitive textures such as sand or grass. Overall, both DIM+Ours and SfM+Ours demonstrate promising performance, highlighting the effectiveness of the proposed volume-based densification method across different initialization strategies.

| Scenes | Strategies | | | |
|---|---|---|---|---|
| | SfM + 3DGS | DIM + 3DGS | SfM + Ours | DIM + Ours |
| Bicycle | 0.246 | 0.272 | 0.230 | **0.226** |
| Flowers | 0.289 | 0.288 | 0.27 | **0.255** |
| Bonsai | 0.145 | 0.177 | **0.128** | 0.153 |
| Garden | 0.153 | 0.128 | **0.101** | 0.118 |
| Room | 0.172 | 0.219 | **0.161** | 0.194 |
| Kitchen | 0.096 | 0.119 | **0.094** | 0.118 |
| Truck | 0.354 | 0.161 | 0.359 | **0.115** |
| Treehill | 0.305 | 0.359 | **0.287** | 0.331 |
| Stump | 0.214 | 0.225 | **0.208** | 0.224 |
| Playroom | 0.247 | 0.304 | **0.234** | 0.289 |
| Dr.Johnson | 0.44 | 0.449 | **0.439** | 0.451 |
| Counter | 0.183 | 0.206 | **0.175** | 0.194 |

Table 2: The effect of different initialization strategies (LPIPS scores).

## 6. Conclusions and perspectives

In this paper, we revisit the densification mechanism involved in the 3D Gaussian Splatting method. Our key observation is that the densification mechanism proposed by Kerbl et al. [4] does not effectively densify regions with relatively low point densities. As a result, the generated ellipsoids tend to be oversized, leading to artifacts such as blurred regions. To address this limitation, our complimentary method introduces additional points based on a volumetric criterion, resulting in more accurate densification. Our approach performs especially well on the MiP-NeRF 360 dataset, with notable improvements in perceptual LPIPS scores. While our method leverages the volume of inertia with a fixed threshold of 0.03 to guide densification, this static parameter may not optimally adapt to scene geometries or levels of detail beyond the considered dataset.

A promising future direction will concern the extension of the proposed densification method to dynamic scenes, where ellipsoids distributions continuously evolve over time. Incorporating volume-based densification criteria and leveraging temporal information from consecutive frames can provide more accurate representations for such dynamic scenes. This extension would significantly enhance the applicability of our approach to real-time rendering and dynamic scene reconstruction tasks.